# Workflow-as-a-Service Cloud Platform and Deployment of Bioinformatics Workflow Applications

*Muhammad H. Hilman, Maria A. Rodriguez, and Rajkumar Buyya*

Cloud Computing and Distributed Systems (CLOUDS) Laboratory

School of Computing and Information Systems, The University of Melbourne, Australia

*Email: hilmanm@student.unimelb.edu.au, {maria.read, rbuyya}@unimelb.edu.au*

Workflow management systems (WMS) support the composition and deployment of workflow-oriented applications in distributed computing environments. They hide the complexity of managing large-scale applications, which includes the controlling data pipelining between tasks, ensuring the application's execution, and orchestrating the distributed computational resources to get a reasonable processing time. With the increasing trends of scientific workflow adoption, the demand to deploy them using a third-party service begins to increase. Workflow-as-a-service (WaaS) is a term representing the platform that serves the users who require to deploy their workflow applications on third-party cloud-managed services. This concept drives the existing WMS technology to evolve towards the development of the WaaS cloud platform. Based on this requirement, we extend CloudBus WMS functionality to handle the workload of multiple workflows and develop the WaaS cloud platform prototype. We implemented the **E**lastic **B**udget-constrained resource **P**rovisioning and **S**cheduling algorithm for **M**ultiple workflows (EBPSM) algorithm that is capable of scheduling multiple workflows and evaluated the platform using two bioinformatics workflows. Our experimental results show that the platform is capable of efficiently handling multiple workflows execution and gaining its purpose to minimize the makespan while meeting the budget.

**Keywords:** Scientific workflow, Cloud computing, Workflow management system, Bioinformatics workflow, Budget-constrained scheduling.

## 1. Introduction

Workflow is a computational model that represents the application tasks and its related flow of data in the form of interconnected nodes. The applications that utilize the workflow model consist of several complexes, large-scale applications, and involve a vast amount of data. Therefore, these workflows are usually deployed in the distributed systems that have massive computational resources such as cluster, grid, and cloud computing environments.

To manage the complexity of executing workflows, its interaction with the users, and its connectivity to the resources in distributed systems, the researchers utilize the toolkit called workflow management system (WMS). The WMS hides the complicated orchestration between those coordinated components. It needs to be noted that the interconnected tasks within a workflow have strict dependencies in which the following tasks can be executed whenever the earlier tasks that become its dependencies have finished their execution. Therefore, the critical responsibility of this WMS includes the management of data movement, the scheduling of tasks and preserving their dependencies, and the provisioning of required computational resources from the external distributed systems.

A conventional WMS is designed to manage the execution of a single workflow application. In this case, a WMS is tailored to a particular workflow application to ensure the efficient execution of the workflow. It is not uncommon for a WMS to be built by a group of researchers to deploy a specific application of their research projects. With the advent of the computational infrastructure and the rising trends of workflow model adoption within the scientific community, there is a demand to provide the execution of workflow as a service. Therefore, there is an idea to elevate the functionality of WMS to provide the service for executing workflows in the clouds called the Workflow-as-a-Service (WaaS) cloud platform.

Developing the WaaS cloud platform means leveraging the WMS functionality and minimizing any specific application-tailored in the component of the system. This challenge arises with several issues related to the resource provisioning and scheduling aspect of the WMS. In this work, we focus on designing the resource provisioning and scheduling module within the existing CloudBus WMS [1] for the WaaS cloud platform development. We modify the scheduling modules to fit into the requirements by building on the capability for scheduling multiple workflows. In summary, the main contributions of this chapter are:

- The development of WaaS cloud platform by extending CloudBus WMS.
- The implementation of EBPSM algorithm that is designed to handle multiple workflows scheduling within WaaS cloud platform.
- The case study to analyse the performance of WaaS cloud platform by deploying bioinformatics workflow applications in real cloud computing environments.

The rest of this chapter is organized as follows. Section 2 reviews works of that are related to our discussion. Section 3 describes the development of WaaS cloud platform and its requirements. Furthermore, Section 4 explains the case studies of executing multiple workflows in WaaS cloud platform. Finally, the Section 5 summarizes the findings and discusses the future directions.

## 2. Related Work

WMS technology has evolved since the era of cluster, grid, and current cloud computing environments. A number of widely used WMS were initially built by groups of multi-disciplinary researchers to deploy the life-science applications of their research projects developed based on the computational workflow model. Each of them has a characteristic tailored to their requirements. However, to the best of our knowledge, the existing WMS systems are not designed for handling multiple workflows execution as it becomes the main requirement for WaaS cloud platform. Therefore, the case study of several prominent WMS is plentiful and worth to be explored further for the development of such a platform. The summary of these characteristics is depicted in Table 1.

**Table 1:** Summary of various WMS features.

| | Main features | ASKALON | Galaxy | HyperFlow | Kepler | Pegasus | Taverna | CloudBus |
|---|---|---|---|---|---|---|---|---|
| Workflow Engine | Service-oriented | ✓ | ✓ | - | ✓ | ✓ | ✓ | ✓ |
| | GUI-supported | ✓ | ✓ | - | ✓ | ✓ | ✓ | ✓ |
| | Provenance-empowered | ✓ | ✓ | ✓ | ✓ | ✓ | | ✓ |
| Distributed Environments | Grid-enabled | ✓ | ✓ | ✓ | ✓ | ✓ | ✓ | ✓ |
| | Cloud-enabled | ✓ | ✓ | ✓ | ✓ | ✓ | ✓ | ✓ |
| | Container-enabled | - | ✓ | ✓ | - | ✓ | - | - |
| | Serverless-enabled | - | - | ✓ | - | - | - | - |

ASKALON [2] is a framework for development and runtime environments for scientific workflows built by a group from The University of Innsbruck, Austria. Along with ASKALON, the group released a novel workflow language standard developed based on the XML called Abstract Workflow Description Language (AWDL) [3]. ASKALON has a tailored implementation of wien2k workflow [4], a material science workflow for performing electronic structure calculations using density functional theory based-on the full-potential augmented plane-wave to be deployed within the Austrian Grid Computing network.

Another project is Galaxy [5], a web-based platform that enables users to share workflow projects and provenance. It connects to myExperiments [6], a social network for sharing the workflow configuration and provenance among the scientific community. It is a prominent WMS and widely used for in silico experiments [7] [8] [9].

A lightweight WMS, HyperFlow [10] is a computational model, programming approach, and also a workflow engine for scientific workflows from AGH University of Science and Technology, Poland. It provides a simple declarative description based on JavaScript. HyperFlow supports the workflow deployment in container-based infrastructures such as docker and Kubernetes clusters. HyperFlow is also able to utilize the serverless architecture for deploying Montage workflow in AWS Lambda and Google Function [11].

Kepler [12] is a workflow management system developed by a collaboration of universities, including UC Davis, UC Santa Barbara, and UC San Diego, United States. It is a WMS that is built on top of the data flow-oriented Ptolemy II system [13] from UC Berkeley. Kepler has been adopted in various scientific projects including the fluid dynamics [14] and computational biology [15]. This WMS provides compatibility to run on different platforms, including Windows, OSX, and Unix systems.

Another project is Pegasus [16], one of the prominent WMS that is widely adopted for projects that make an essential breakthrough to scientific discovery from The University of Southern California, United States. Pegasus runs the workflows on top of HTCondor [17] and supports the deployment across several distributed systems, including grid, cloud, and container-based environments. The Pegasus WMS has a contribution to the LIGO projects involved in the gravitational wave detection [18].

Furthermore, there is also Taverna [19], a workflow management system from The University of Manchester that as recently accepted under the Apache Incubator project. Taverna is designed to enable various deployment models from the standalone, server-based, portal, clusters, grids, to the cloud environments. Taverna has been used in various in silico bioinformatics projects, including several novel Metabolomics research [20] [21].

Finally, the CloudBus WMS [1], a cloud-enabled WMS from The University of Melbourne, is the center of discussion in this chapter. Its functionality evolves to support the development of the WaaS cloud platform.

## 3. Prototype of WaaS Cloud Platform

In this section, we discuss a brief development of the CloudBus WMS and the WaaS cloud platform development. We describe the evolving functionality of CloudBus WMS in its first release to handle the deployment in the grid computing environment up to the latest version that provides the cloud-enabled functionality to give an overview of how the distributed systems trend changes how the WMS works. Furthermore, we present the extension related to the scheduler component of this WMS to support the development of the WaaS cloud platform.

### 3.1. CloudBus Workflow Management System

The earliest version of the WMS from the CLOUDS lab was designed for grid computing environments under the name of GridBus Workflow Enactment Engine in 2008. The core engine in this WMS was called a workflow enactment engine that orchestrated the whole workflow execution. The engine interacts with users through the portal that manages workflow composition and execution planning. This engine also equipped with the ability to interact with grid computing environments through the grid resource discovery to find the possible grid computational infrastructure, the dispatcher that sends the tasks to the grids for the execution, and the data movement to manage data transfer in and out through HTTP and GridFTP protocols. The Gridbus Workflow Enactment Engine was tested and evaluated using a case study of fMRI data analysis in the medical area. The architectural reference to this Gridbus Workflow Engine and its case study can be referred to the paper by Yu and Buyya [22].

The second version of the GridBus Workflow Enactment Engine was released in 2011, built with plugin support for deployment in cloud computing environments. In this version, the engine is equipped with the components that enable it to utilize several types of external computational resources, including grid and cloud environments. Therefore, it was renamed to CloudBus Workflow Engine. In addition to this functionality, the CloudBus Workflow Engine was tested and evaluated for scientific workflow execution on top of the Aneka Cloud Enterprise platform [23] and Amazon Elastic Compute Cloud (EC2) using a case study of evolutionary multiobjective optimization technique based on a genetic algorithm. We suggested that readers refer to the architectural design and case study implementation published by Pandey et al. [24].

The latest release of the CloudBus Workflow Engine in 2016 was the implementation of a comprehensive cloud-enabled functionality that allows the engine to lease the computational resources dynamically from the IaaS cloud providers. This version introduces a Cloud Resource Manager module that enables the platform to manage the resources (i.e., Virtual Machines) from several IaaS cloud providers related to its automated provisioning, integrating to the resource pool, and terminating the VMs based on the periodic scanning of the implemented algorithm. Along with the dynamic functionality of cloud resources management, the WMS is also equipped with a dynamic algorithm to schedule workflows which able to estimate the tasks' runtime based on the historical data from the previous workflows' execution. This version is known as the CloudBus Workflow Management System (WMS). The architectural reference and its case study on Astronomical application Montage can be referred to the paper by Rodriguez and Buyya [1].

### 3.2. WaaS Cloud Platform Development

The CloudBus WMS is continuously adapting to the trends of the distributed systems infrastructures from cluster, grid, to the cloud environments. With the increasing popularity of the computational workflow model across scientific fields, we extend the CloudBus WMS to serve as a platform that provides the execution of workflow as a service. Therefore, we design the reference to the WaaS cloud platform based on the structure of CloudBus WMS. Five entities compose the WaaS cloud platform, they are portal, engine, monitoring service, historical database, and plugins to connect to distributed computing environments. This structure is similar to the previous CloudBus WMS architecture. The architectural reference for the platform can be seen in Figure. 1.

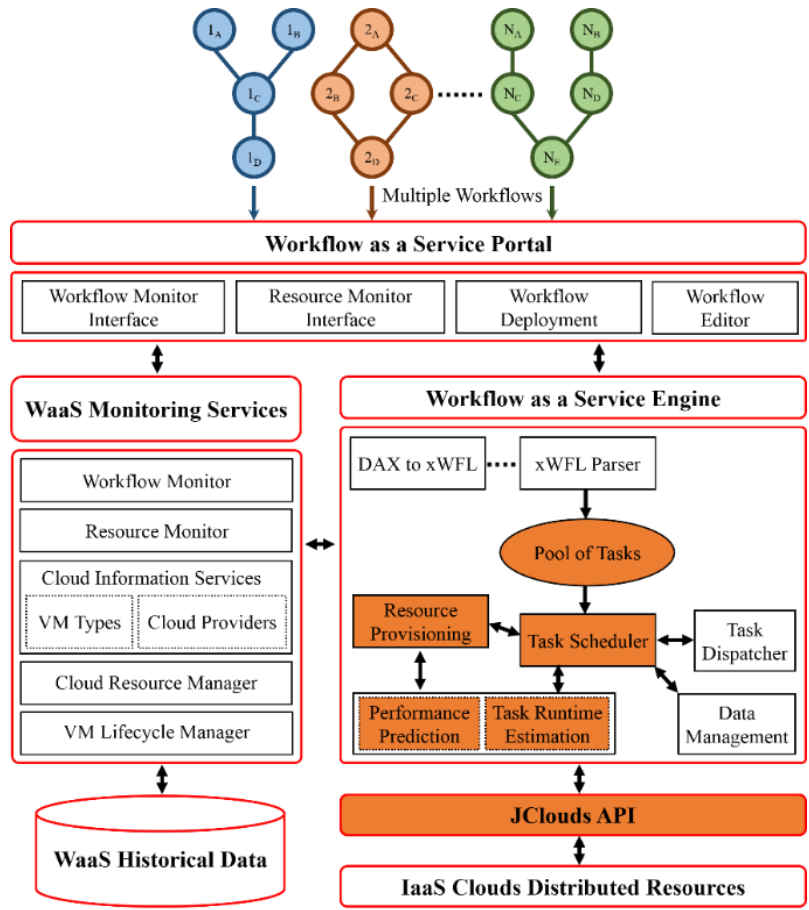

**Figure 1:** Architectural reference on the WaaS cloud platform.

**Portal:** an entity that is responsible for bridging the WaaS cloud platform to the users. The portal serves as the user interface in which users can submit the job, including composing, editing, and defining the workflow QoS requirements. It interacts with the engine to pass on the submitted workflows for scheduling. It also interacts with the monitoring service so that the users can monitor the progress of the workflows' deployment. Finally, the engine sends back the output data after it finished the execution through this entity. The change from the previous CloudBus WMS functionality is the capability of the portal to handle the workload of multiple workflows.

**Monitoring Service:** an entity that is responsible for monitoring the workflow execution and resources running within the WaaS cloud platform that is provisioned from the clouds. Five components in this entity are the *Workflow Monitor* that tracks the execution of the jobs, the *Resource Monitor* which tracks the VMs running in the platform, the *Cloud Information Services* that discover the available VM types and images of the IaaS clouds profile, the *Cloud Resource Manager* that manages the provisioning of cloud resources, and the *VM Lifecycle Manager* which keeps tracking the VMs' status before deciding to terminate them.

This entity interacts with the portal to provide the monitoring information of workflows' execution. On the other hand, it also interacts with the engine to deliver the status of job execution for scheduling purposes and the state of the computational resource availability. We changed the provisioning algorithm, which is managed by the cloud resource manager and the VM lifecycle manager, based on the EBPSM algorithm. Both the cloud resource manager and the VM lifecycle manager control the VMs provisioning by keeping track of the idle status of each VM. They will be terminated if the idle time exceeded the *threshold$_{idle}$*. This provisioning algorithm is depicted in Algorithm 1. Finally, this entity saves the historical data of tasks' execution into the historical database based on an HSQL database where the information is used to estimate the tasks' runtime.

**Engine:** an entity that is responsible for the orchestration of the whole execution of workflows. This entity interacts with the other objects of the WaaS cloud platform, including the third-party services outside the platform. Moreover, it takes the workflows' job from the portal and manages the execution of tasks. The scheduler that is part of this entity schedules each task from different workflows and allocates them to the available resources maintained by the monitoring service. It also sends the request to the plugins, JClouds API, for provisioning new resources if there is no available idle VMs to reuse.

*Task scheduler*, the core of the engine, is modified to adapt to the EBPSM algorithm that manages the scheduling of multiple workflows. Within the task scheduler, there is a component called the *WorkflowCoordinator* that creates the *Task Manager(s)* responsible for scheduling each task from the pool of tasks. To manage the arriving tasks from the portal, we create a new class *WorkflowPoolManager* responsible for periodically releasing ready tasks for scheduling and keeping track of the ownership of each task.

*Prediction* component within the task scheduler is responsible for estimating the runtime of the task, which becomes a pre-requisite of the scheduling. We modify the *PredictRuntime* component to be capable of building an online incremental learning model. This learning model is a new approach for estimating the runtime for scientific workflows implemented in the WaaS cloud platform. While previously, this module utilizes the statistical analysis approach.

**Historical database:** an HSQL database used to store the historical data of tasks' execution. The information, then, is used to estimate the tasks' runtime. In this platform, we add the submission time variables to the database, since this information is used to build the prediction model to estimate the runtime.

**Plugins:** a JClouds API responsible for connecting the WaaS cloud platform to third party computational resources. Currently, the platform can connect to several cloud providers, including Amazon Elastic Compute Cloud (EC2), Google Cloud Engine, Windows Azure, and OpenStack-based NeCTAR clouds. It sends the request to provision and terminates resources from the cloud providers.

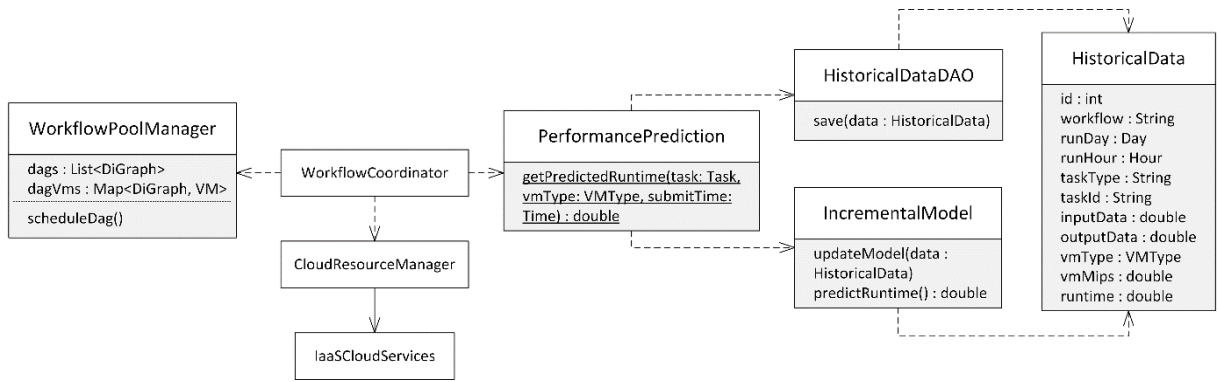

**Figure 2:** Class diagram reference on the scheduler extension of the WaaS cloud platform.

Finally, the modified components within the WaaS cloud platform from the previous version of the CloudBus WMS are marked with the red-filled diagram in Figure. 1 and the class diagram reference to the WaaS cloud platform scheduler extension are depicted in Figure. 2.

### 3.3. Implementation of Multiple Workflows Scheduling Algorithm

**E**lastic **B**udget-constrained resource **P**rovisioning and **S**cheduling algorithm for **M**ultiple workflows is a dynamic heuristic algorithm designed for WaaS cloud platform. The algorithm was designed to schedule tasks from multiple workflows driven by the budget to minimize the makespan. EBPSM distributes the budget to each of its tasks in the first step, and then, it manages the tasks from different workflows to schedule based on its readiness to run (i.e., parents' tasks finished the execution).

Furthermore, the algorithm looks for idle resources that can finish the tasks as fast as possible without violating its assigned budget. This algorithm enforces the reuse of already provisioned resources (i.e., virtual machines) and sharing them between tasks from different workflows. This policy was endorsed to handle the uncertainties in the clouds, including VM performance variability, VM provisioning, and deprovisioning delays, and the network-related overhead that incurs within the environments. Whenever a task finishes, the algorithm redistributes the budget for the task's children based on the actual cost. In this way, the uncertainties, as mentioned earlier from cloud computing environments, can be further mitigated before creating a snowball effect for the following tasks.

The scheduling phase of the EBPSM algorithm was mainly implemented in the task scheduler, a part of the engine. The *WorkflowPoolManager* class receives the workflows' jobs and distributes the budget to the tasks as described in Algorithm 2. It keeps track of the workflows' tasks before placing the ready tasks on the priority queue based on the ascending Earliest Finish Time (EFT). Then, the *WorkflowCoordinator* creates a task manager for each task that is pooled from the queue. In the resource provisioning phase, the task scheduler interacts with the cloud resource manager in the monitoring resource to get the information of the available VMs. The task scheduler sends the request to provision a new VM if there are no VMs available to reuse. The implementation of this phase involving several modules from different components of the WaaS cloud platform. The detail of this scheduling is depicted in Algorithm 3.

The post-scheduling of a task ensures budget preservation by calculating the actual cost and redistributing the workflows' budget. This functionality was implemented in the task scheduler with additional information related to the clouds from the cloud information service, which maintains the cloud profile such as the VM types, and the cost of the billing period. The detail of the budget re-distribution procedure is described in Algorithm 4.

In this work, we implemented a version of the EBPSM algorithm without the container. We did not need the container-enabled version as we only used bioinformatics workflow applications that did not have conflicting software dependencies and libraries. The enablement for microservices and serverless-supported WaaS cloud platform is left for further development. For more details on the EBPSM versions and their budget distribution strategies, we suggested the readers to refer to the papers by Hilman et al. [25] [26].

**Algorithm 1:** Resource Provisioning
**procedure** manageResource
    $VM_{idle}$ = all leased VMs that are idle
    $threshold_{idle}$ = idle time threshold
    **for** each $vm_{idle} \in VM_{idle}$ **do**
        $t_{idle}$ = idle time of $vm$
        **if** $t_{idle} \geq threshold_{idle}$ **then**
            terminate $vm_{idle}$
        **end if**
    **end for**
**end procedure**

---

**Algorithm 2:** Budget Distribution
$\beta$ = workflow's budget
$T$ = set of tasks in the workflow
**procedure** distributeBudget ($\beta$, $T$)
    $S$ = tasks' estimated execution order
    $l$ = tasks' level in the workflow
    **for** each $task\ t \in T$ **do**
        *allocateLevel(t, l)*
        *initiateBudget(0, t)*
    **end for**
    **for** each level $l$ **do**
        $T_l$ = set of all tasks in level $l$
        sort $T_l$ based on ascending EFT
        *put($T_l$, S)*
    **end for**
    **while** $\beta > 0$ **do**
        $t = S.poll$
        $C_{vmt}^{t}$ = cost of task $t$ in $vmt$
        $vmt$ = chosen VM type
        *allocateBudget($C_{vmt}^{t}$, t)*
        $\beta = \beta - C_{vmt}^{t}$
    **end while**
**end procedure**

---

**Algorithm 3:** Scheduling
$q$ = queueing tasks for scheduling
**procedure** scheduledQueuedTasks($q$)
    sort $q$ by ascending EFT
    **while** $q$ is not empty **do**
        $t = q.poll$
        $vm = null$
        **if** there are idle VMs **then**
            $VM_{idle}$ = set of all idle VMs
            $vm = vm \in VM_{idle}$ that can finish $t$ within $t.budget$ with the fastest execution time
        **else**
            $vmt$ = fastest VM type within $t.budget$
            $vm = provisionVM(vmt)$
        **end if**
        *scheduleTask(t, vm)*
    **end for**
**end procedure**

---

**Algorithm 4:** Budget Update
**procedure** updateBudget($T$)
    $t_f$ = completed task
    $T_u$ = set of unscheduled $t \in T$
    $\beta_u$ = total sum of t.budget, where $t \in T_u$
    $sb$ = spare budget
    **if** $C_{vmt}^{t_f} \leq (t_f.budget + sb)$ **then**
        $sb = (t_f.budget + sb) - C_{vmt}^{t_f}$
        $\beta_u = \beta_u + sb$
    **else**
        $debt = C_{vmt}^{t_f} - (t_f.budget + sb)$
        $\beta_u = \beta_u - debt$
    **end if**
    *distributeBudget($\beta_u$, $T_u$)*
**end procedure**

## 4. Case Studies and Performance Evaluation

In this section, we present the case study of multiple workflows execution within a WaaS cloud platform prototype. We address the workload of bioinformatics workflows and its preparation for the execution. Furthermore, we also describe the technical infrastructure and its experimental design to deploy the platform and present the results from the experiment.

## 4.1. Bioinformatics Applications Workload

Many bioinformatics cases have adopted the workflow model for managing its scientific applications. An example is myExperiments [6] that has a broader scope to connect various bioinformatics workflows users. This social network for scientists who utilize the workflows for managing their experiments, stores almost four thousand workflows software, configurations, and datasets with more than ten thousand members. We explored two prominent bioinformatics workflows in the area of genomics analysis [27] and drug discovery [28] for the case study.

### *4.1.1. Identifying Mutational Overlapping Genes*

The first bioinformatics case was based on the 1000 Genomes Project[1], an international collaboration project to build a human genetic variation catalogue. Specifically, we used an existing 1000 Genome workflow[2] to identify overlapping mutations in humans' genes. The overlapping mutations were statistically calculated in a rigorous way to provide an analysis of possible disease-related mutations across human populations based on their genomics properties. This project has an impact on evolutionary biology. Examples include a project related to the discovery of full genealogical histories of DNA sequences [29].

The workflow consists of five tasks that have different computational requirements [30]. They are *individuals*, *individuals_merge*, *sifting*, *mutations_overlap*, and *frequency*. *Individuals* performs data fetching and parsing of the 1000 genome project data that listed all Single Nucleotide Polymorphism (SNPs) variation in the chromosome. This activity involves a lot of I/O reading and writing system call. *Individuals_merge* showed similar properties, as it was a merging of *individuals* outputs that calculate different parts of chromosomes data. Furthermore, *sifting* calculates the SIFT scores of all SNPs variants. This task has a very short runtime.

---

[1] http://www.internationalgenome.org/about
[2] https://github.com/pegasus-isi/1000genome-workflow

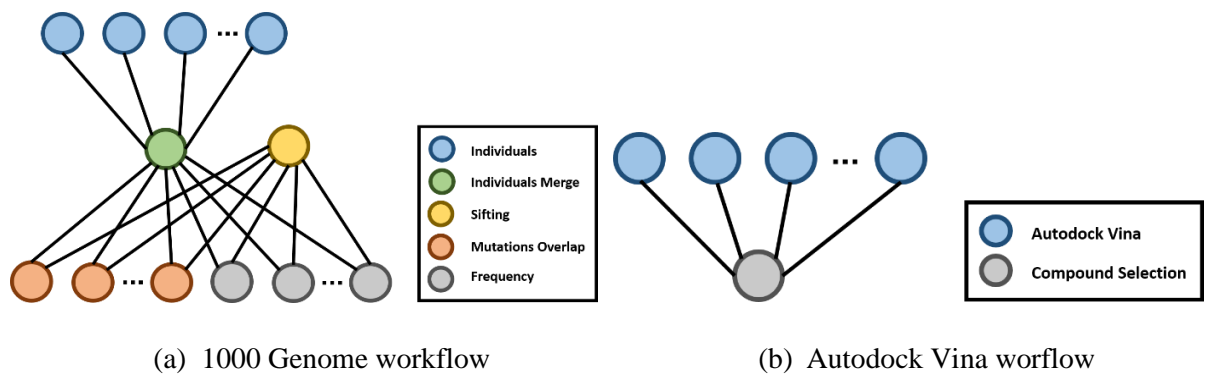

(a) 1000 Genome workflow      (b) Autodock Vina worflow

**Figure 3:** Bioinformatics workflow applications

Finally, *mutations_overlap* calculates the overlapping mutations genes between a pair of individuals while *frequency* calculates the total frequency of overlapping mutations genes between several random individuals.

The 1000 Genome workflow takes two inputs, the chromosome data and its haplotype estimation (i.e., phasing) using the shapeit method. The entry tasks were *individuals*, which extract each individual from chromosome data, and *sifting* that calculates the SIFT scores from the phasing data. Furthermore, in the next level, *individuals_merge* merged all output from *individuals* and then, its output along with the *sifting* output becomes the input for the exit tasks of *mutation_overlap* and *frequency*. For our study, we analyzed the data corresponding to two chromosomes (chr21 and chr22) across five populations: African (AFR), Mixed American (AMR), East Asian (EAS), European (EUR), and South Asian (SAS). Furthermore, the structure of the workflow is shown in Figure. 3a.

### 4.1.2. Virtual Screening for Drug Discovery

The second bioinformatics case used in this study was the virtual screening workflow. Virtual screening is a novel methodology that utilized several computational tools to screen a large number of molecules' libraries for possible drug candidates [31]. In simple terms, this (part of) drug discovery process involves two types of molecules, target receptors, and ligands that would become the candidates of drugs based on its binding affinity to the target receptor. This technique rises in popularity as the in-silico infrastructure and information technology are getting better. The virtual screening saves many resources of scientists for in-vitro and in-vivo that require wet-lab experiments.

There are two main approaches in carrying out the virtual screening, ligand-based, and receptor-based virtual screening [32]. The ligand-based virtual screening relies on the similarity matching of ligands' libraries to the already known active ligand(s) properties. This activity is computationally cheaper than the other approach, as it depends only on the computation of the features of the molecules. On the other hand, the receptor-based virtual screening requires the calculation for both of the target receptors and the ligands to evaluate the possible interaction between them in a very intensive simulation and modelling. However, since the error rate of ligand-based virtual screening is relatively higher than the structure-based, this approach is applied as a filter step when the number of ligands involved in the experiments is quite high.

In this study, we used a virtual screening workflow using AutoDock Vina [33], a molecular docking application for structure-based virtual screening. In particular, we took a virtual screening case of one receptor and ligands with various sizes and search spaces of the docking box taken from the Open Science Grid Project developed by the Pegasus group[3]. The receptor-ligand docking tasks in this workflow can be executed in parallel as in the bag of the tasks application model. Moreover, AutoDock Vina is a CPU-intensive application that can utilize the multi-CPU available in a machine to speed up the molecular docking execution. Therefore, two-level parallelism can be achieved to speed up the workflows, the parallel execution of several receptor-ligand docking tasks on different machines, and the multi-CPU parallel execution of a docking task within a machine. The structure of the virtual screening workflows is depicted in Figure. 3b.

---

[3] https://github.com/pegasus-isi/AutoDock-Vina-Workflow

**Table 2:** Various budgets used in evaluation

| Name | $\beta_1$ | $\beta_2$ | $\beta_3$ | $\beta_4$ |
|---|---|---|---|---|
| 1000 Genome Workflow | | | | |
| chr21 | $0.1 | $0.25 | $0.45 | $0.65 |
| chr22 | $0.1 | $0.25 | $0.45 | $0.65 |
| Virtual Screening Workflow | | | | |
| vina01 | $0.05 | $0.15 | $0.25 | $0.35 |
| vina02 | $0.01 | $0.04 | $0.06 | $0.08 |

### 4.1.3. Workload Preparation

The Pegasus group has developed the tools to generate both the 1000 Genome and Virtual Screening workflow based on the XML format. We converted the DAG generated from the tools into the xWFL, the format used by the WaaS cloud platform. Based on this converted-DAG, we prepared two versions of the 1000 Genome workflows, which take two different chromosomes of chr21 and chr22 as input. Furthermore, we created two types of workflows that take as input two different sets of 7 ligands molecules for Virtual Screening.

We installed five applications for the 1000 Genome workflow in a custom VM image for the worker nodes. These applications are based on the Mutation_Sets project[4] and are available in the 1000 Genome workflow project. It needs to be noted that the *mutation_overlap* and *frequency* tasks were python-based applications and have a dependency to the *python-numpy* and *python-matplotlib* modules. On the other hand, the only application that needs to be installed for the Virtual Screening workflow was AutoDock Vina, which can be installed without any conflicting dependencies with the other workflow applications. Therefore, in this scenario, we did not encounter the conflicting dependencies problem.

---

[4] https://github.com/rosafilgueira/Mutation\_Sets

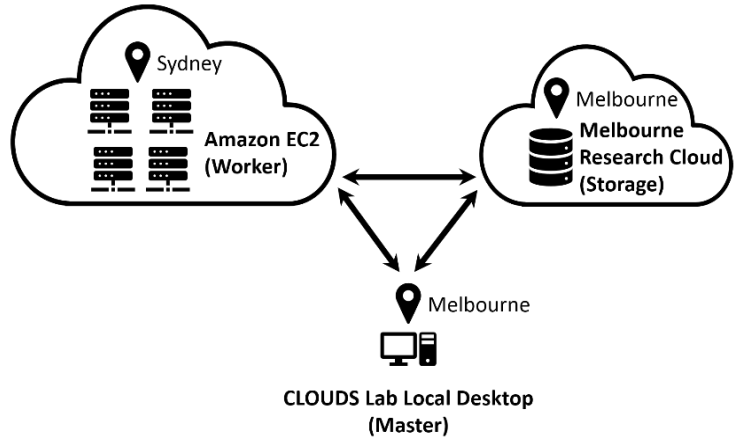

**Figure 4:** Architectural reference on the WaaS cloud platform nodes deployment

We composed a workload that consists of 20 workflows with the types as mentioned earlier of applications that were randomly selected based on a uniform distribution. We also modelled four different arrival rates of those workflows based on a Poisson distribution from 0.5 workflows per minute (wf/m), which represents the infrequent requests, up to 12 wf/m that reflect the busiest hours. Each workflow was assigned a sufficient budget based on our initial deployment observation. We defined four different budgets for each workflow from $β_1$ to $β_4$, which represents the minimum to the maximum willingness of users to spend for particular workflows' execution. These budgets can be seen in Table 2.

### 4.2. Experimental Infrastructure Setup

Three components need to be deployed to ensure the running of the WaaS cloud platform. The first is the master node containing the core of the workflow engine. This master node is the component that manages the lifecycle of workflows execution and responsible for the automated orchestration between every element within the platform. The second component is a storage node which stores all the data involved in the execution of the workflows. This storage manages the intermediate data produced between parents and children tasks' execution and acts as a central repository for the WaaS cloud platform. Finally, the worker node(s) is the front runner(s) to execute the workflows' tasks submitted into the platform. The worker node(s) provisioning and lifespans are controlled based on the scheduling algorithms implemented in the core of the workflow engine.

**Table 3:** Configuration of virtual machines used in evaluation

| Name | vCPU | Memory | Price per second |
|---|---|---|---|
| CLOUDS Lab Local Desktop | | | |
| Master Node | 4 | 8192 MB | N/A |
| Melbourne Research Cloud | | | |
| Storage Node | 1 | 4096 MB | N/A |
| Amazon EC2 | | | |
| Worker Node | | | |
| *t2.micro* | 1 | 1024 MB | $0.0000041 |
| *t2.small* | 1 | 2048 MB | $0.0000082 |
| *t2.medium* | 2 | 4096 MB | $0.0000164 |
| *t2.large* | 2 | 8192 MB | $0.0000382 |

For this experiment, we arranged these components on virtual machines with different configurations and setup. The master node was installed on Ubuntu 14.04. 6 LTS virtual machine running in a local HP Laptop with Intel(R) Core(TM) i7-56000 CPU @ 2.60 GHz processor and 16.0 GB RAM. This virtual machine was launched using VMWare Workstation 15 player with 8.0 GB RAM and 60.0 GB hard disk storage. Moreover, we deployed the storage node on a cloud instance provided by The Melbourne Research Cloud[5] located in the *melbourne-qh2-uom* availability zone. This virtual machine was installed Ubuntu 14.04.6 LTS operating systems based on the *uom. general.1c4g* flavour with 1 vCPU, 4 GB RAM, and an additional 500 GB hard disk storage.

Furthermore, the worker node(s) were dynamically provisioned on Amazon Elastic Compute Cloud (EC2) Asia Pacific Sydney region using a custom prepared VM image equipped with the necessary software, dependencies, and libraries for executing 1000 Genome and Virtual Screening workflows. We used four different types and configurations for the worker nodes based on the family of T2 instances. The T2 instances family equipped with the high-frequency processors and have a balance of compute, memory, and network resources. Finally, the architectural reference for the nodes' deployment and its configuration are depicted in Figure. 4 and Table. 3 respectively.

---

[5] https://research.unimelb.edu.au/infrastructure/research-computing-services/services/research-cloud

### 4.3. Results and Analysis

In this section, we present the comparison of EBPSM and First Come First Serve (FCFS) algorithm, as the default scheduler, in a single workflow and homogeneous settings to ensure the fair evaluation. Then, it was followed by a thorough analysis of the EBPSM performance on a workload of multiple workflows in a heterogeneous environment represented by different arrival rates of workflows to the WaaS cloud platform.

#### *4.3.1. More Cost to Gain Faster Execution*

The purpose of this particular experiment is to evaluate our proposed EBPSM algorithm for the WaaS platform compared to the default scheduler of the CloudBus WMS. This default scheduler algorithm did not rely on an estimate of tasks' runtime. It scheduled each task based on the first-come, first-served policy into a dedicated resource (i.e., VM) and terminated the resource when the particular task has finished the execution. Furthermore, this default scheduler was not equipped with the capability to select the resources in heterogeneous environments. Therefore, it only works for homogeneous cluster settings (i.e., clusters of one VM type only). Then, to have a fair comparison to the default scheduler, we modified the EBPSM algorithm to work for a single workflow in a homogeneous environment. We removed the module that enables EBPSM to select the fastest resources based on the task's sub-budget and let the algorithm provision a new VM if there are no idle VMs available to reuse, which means hiding the budget-driven ability of the algorithm.

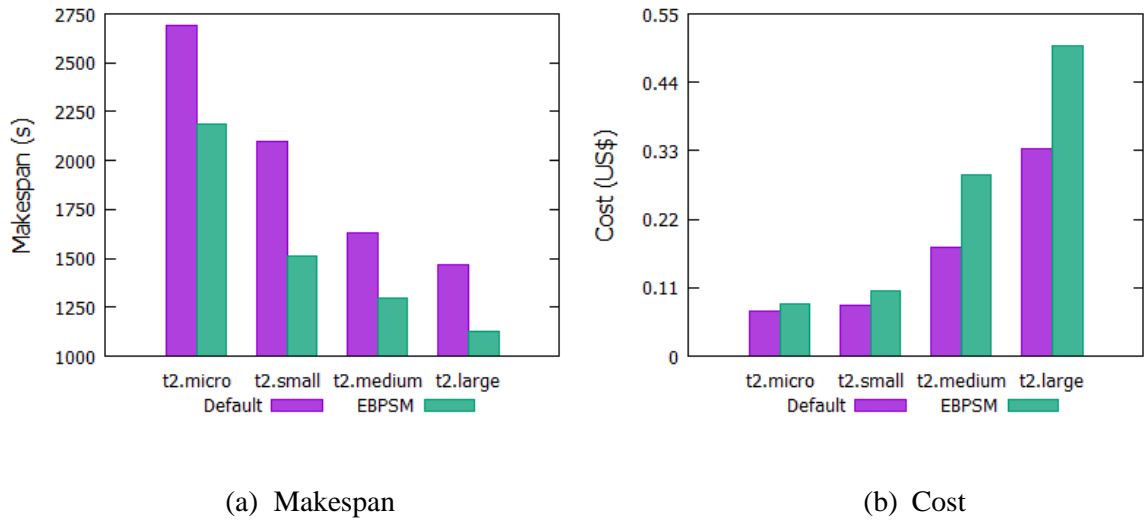

(a) Makespan                  (b) Cost

**Figure 5:** Makespan and cost of chr22 workflow on homogeneous environment

In Figure. 5a, we can see that the homogeneous version of EBPSM was superior to the default scheduler on all scenarios. In this experiment, the default scheduler provisioned 26 VMs for each situation, while EBPSM only leased 14 VMs. In this case, we argue that the delays in initiating the VMs, which include the provisioning delay and delays in configuring the VM into the WaaS platform, have a significant impact on the total makespan. Therefore, the EBPSM can gain an average speedup of 1.3x faster compared to the default scheduler. However, this enhancement comes with a consequence of additional monetary cost.

Figure. 5b showed that there is an increase in monetary cost for executing the workflows. The EBPSM algorithm lets the idle VM to active for a certain period before being terminated, hoping that the next ready tasks would reuse it. This approach produced a higher cost compared to the immediate resource termination of the default scheduler approach. The average increase was 40% higher than the default scheduler. Is it worth to spend 40% more cost to gain 1.3x faster makespan? Further evaluation, such as Pareto analysis, needs to be done. However, more rapid responses to events such as modelling the storm, tsunami, and bush fires in the emergency disaster situation, or predicting the cell location for critical surgery are undoubtedly worth more resources to be spent.

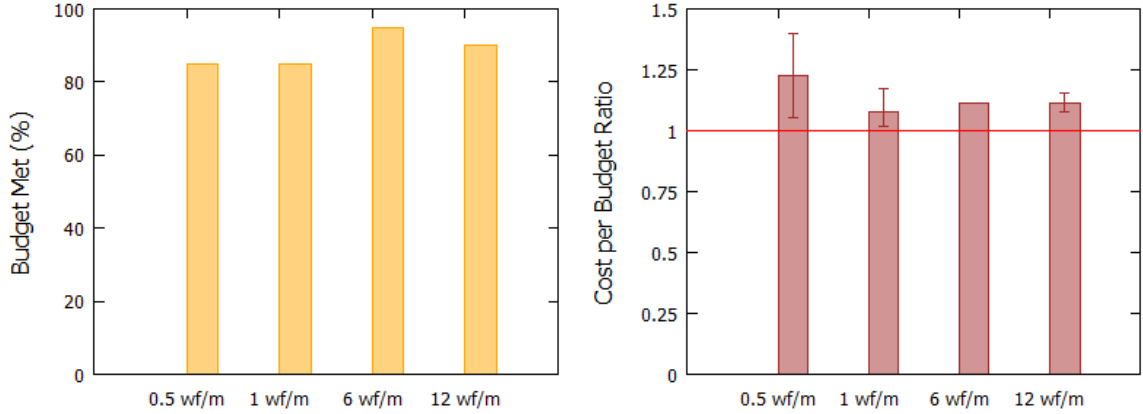

(a) Percentage of budget met  (b) Cost per budget ratio

**Figure 6:** Cost and budget analysis on workload with different arrival rate

*4.3.2. Budget Met Analysis*

To evaluate the budget-constrained multiple workflows deployment, we analyzed the performance of the EBPSM against its primary objective, meeting the budget. Two metrics were used in this analysis, the number of successful cases in meeting the budget, and the cost per budget ratio for any failed ones.

In this experiment, we observed the EBPSM performance in various arrival rate scenarios to see if this algorithm can handle the workload both in peak and non-peak hours. Figure. 6a showed that in the non-peak hours, the EBPSM could achieve 85% of the budget met while in the busier environment, this percentage increases up to 95%. In the peak-hours, there are more VMs to reuse and less idle time that makes the platform more efficient. However, it needs to be noted that there might exist some variability in the Amazon Elastic Compute Cloud (EC2) performance that might impact the results. Thus, the graphs did not show a linear convergence. Nevertheless, 85% of the budget-met percentage showed satisfactory performance.

The result of failed cases is depicted in Figure. 6b. From this figure, we can confirm the superiority of EBPSM for the peak-hours scenarios. The violation of the user-defined budget was not more than 15% in the peak-hours while the number increases up to 40% can be observed in the non-peak hours' settings. On average, the budget violation was never higher than 14% for all arrival rate schemes. Still and all, this violation was inevitable due to the performance variation of the Amazon Elastic Compute Cloud (EC2) resources.

**Table 4:** Comparison of chr22 workflow in two environments

| Name | Makespan (s) | | Cost ($) | |
| --- | --- | --- | --- | --- |
| | Minimum | Maximum | Minimum | Maximum |
| Single – Homogeneous | 2187 | 1125 | 0.084 | 0.499 |
| Multiple – Heterogeneous | 1819 | 1013 | 0.062 | 0.471 |

### 4.3.3. *Makespan Evaluation*

It is essential to analyze the impact of scheduling multiple workflows on each of the workflows' makespan. We need to know whether sharing the resources between various users with different workflows is worth it and more efficient compared to a dedicated resource scenario in deploying the workflows. Before we discussed further, let us revisit the Figure. 5a, which showed the result of a single 1000 Genome (chr22) workflow execution in a homogeneous environment. Then, we compared it to the Figure. 7b that presented the result for the same 1000 Genome (chr22) workflow in multiple workflows scenario and heterogeneous environment. If we zoomed-in to the two figures, we could observe that EBPSM can further reduce both the makespan and the cost for the workflow in the latter scenario. We extracted these details of both scenarios into Table 4.

Let us continue the discussion for the makespan analysis. Figure. 7a, 7b, 8a, and 8b depicted the makespan results for 1000 Genome (chr21, chr22) and Virtual Screening (vina01, vina02) respectively. If we glanced, there was no linear pattern showing the improvement of EBPSM performance over the different arrival rates of workflows. Nevertheless, if we observed further and split the view into two (i.e., peak hours and non-peak hours), we can see that the EBPSM, in general, produced better results for the peak-hour scenarios except for some outlier from 1000 Genome (chr22) and Virtual Screening (vina01) workflows. We thought that this might be caused by the number of experiments and the size of the workload. This is an important note to be taken as, due to the limited resources, we could not deploy workload with the scale of hundreds, even thousands of workflows.

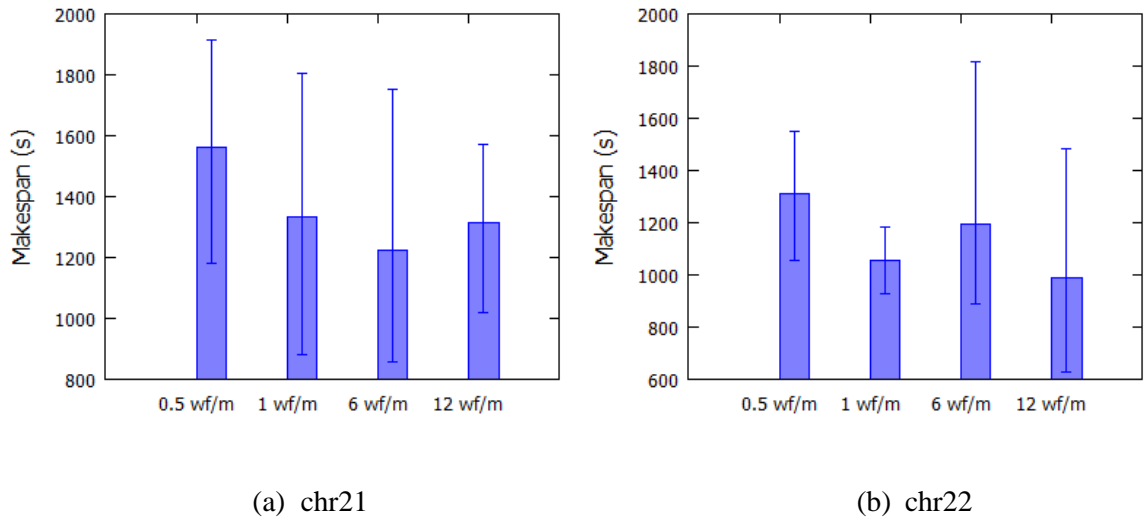

(a) chr21

(b) chr22

**Figure 7:** Makespan of 1000 Genome workflows on workload with different arrival rate

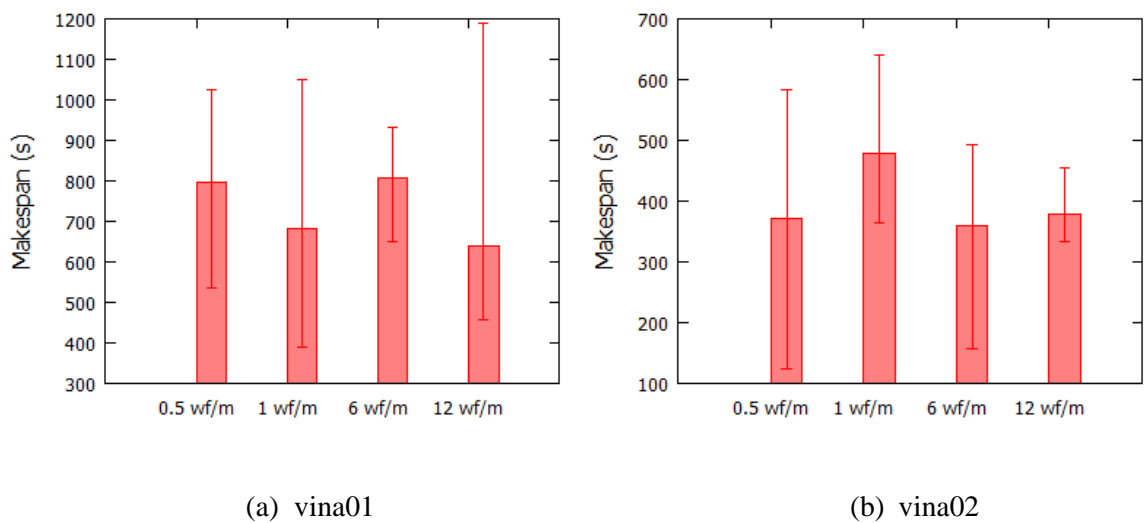

(a) vina01

(b) vina02

**Figure 8:** Makespan of virtual screening workflows on workload with different arrival rate

*4.3.4. VM Utilization Analysis*

Finally, the last aspect to be evaluated regarding the EBPSM performance was VM utilization. It was the most important thing to be pointed out when discussing the policy of sharing and reusing computational resources. In Figure. 9a, we can see the increasing trend in VM utilization percentage along with the arrival rate of workflows on the platform. The average utilization upsurge for each scenario was 4%. The minimum utilization rate was 20% produced by the 0.5 wf/m scenario and the maximum of 36% for the 12 wf/m scenario.

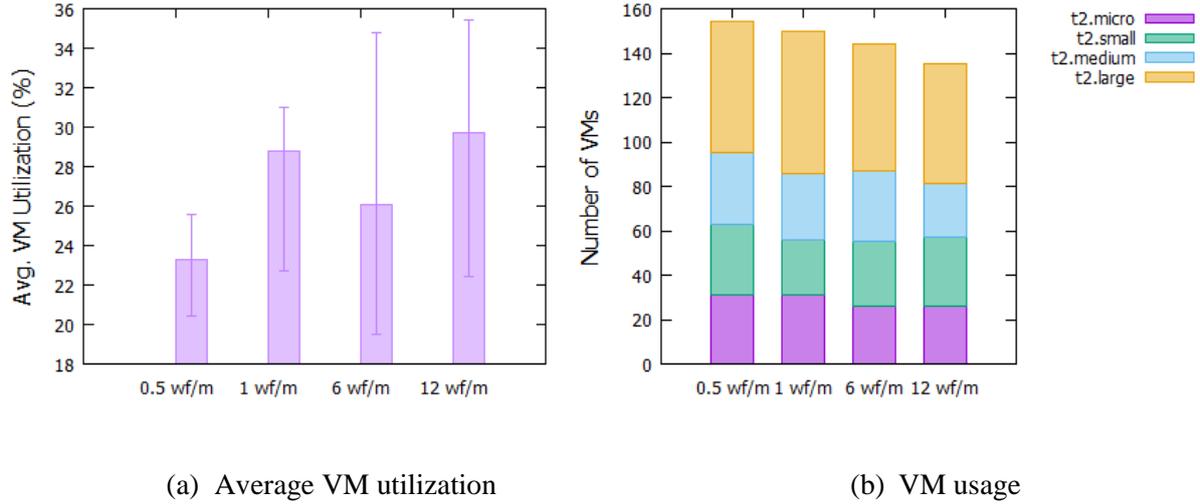

(a) Average VM utilization    (b) VM usage

**Figure 9:** Average VM utilization and VM usage on workload with different arrival rate

We argue that the VM utilization rate had a connection to the number of VMs used during the execution. Figure. 9b depicted the number of VMs used in this experiment. We can observe that the overall number of VMs was declining along with the arrival rate of workflows. The average number of decreases was 20% for all VM types. The lowest drop was for the *t2.large* by 15%, and the highest drop was for the *t2.medium* by 25%. Meanwhile, the *t2.small* decreased by 22% and *t2.micro* by 16% respectively. The EBPSM algorithm always preferred to the fastest VM type and re-calculate and redistribute the budget after each task finished execution. Hence, in this case, the exit tasks might use more VMs of the cheapest type if the budget has been used up by the earlier tasks. Therefore, *t2.large* as the fastest VM type along with *t2.micro* as the cheapest would always be preferred compared to the other VM type.

From this experiment, we concluded that in the WaaS cloud platform where the number of workflows involved is high, the scheduling algorithm must be able to maintain the low number of VMs being provisioned. Any additional VM leased means the higher possibility of incurring more delays related to the provisioning, initiating, and configuring the VMs before being allocated for executing the abundance of tasks.

## 5. Conclusions and Future Work

The workflow management systems (WMS) have a crucial responsibility in executing scientific workflows. It manages the complicated orchestration process in scheduling the workflows and provisioning the required computational resources during the execution of scientific workflows. With the increasing trends of outsourcing computational power to third party cloud providers, there is a consideration to escalate the standalone execution of scientific workflows to the platform that provides the particular service. In this case, there is an emerging concept of a Workflow-as-a-Service (WaaS), extending the conventional WMS functionality to ensure the execution of scientific workflows as a utility service in a WaaS cloud platform.

In this work, we extended the CloudBus WMS by modifying several components for it capable of scheduling multiple workflows to develop the WaaS cloud platform. We implemented the EBPSM algorithm, budget-constrained scheduling algorithm designed for the WaaS platform that is capable of minimizing the makespan while meeting the budget. Furthermore, we evaluated the system prototype using two bioinformatics workflows applications with various scenarios. The experiment results demonstrate that the WaaS cloud platform, along with the EBPSM algorithm, is capable of executing a workload of multiple bioinformatics workflows.

As this work primarily focused on designing the WaaS scheduler functionality, further development of the WaaS cloud platform would be focused on developing the WaaS portal. It is the interface that connects the platform with the users. In this case, the users are expected to be able to compose and define their workflow's job, submit the job and the data needed, monitor the execution, retrieving the output from the workflow's execution. Finalizing the server-based functionality is another to-do list so that the WaaS cloud platform can act as a fully functional service platform in the clouds.

Finally, we plan to enable the WaaS cloud platform for deploying workflows on microservices technology such as container technology, serverless computing, and unikernels system to accommodate the rising demand of the Internet of Things (IoT) workflows. This IoT demand is increasing along with the shifting from centralized infrastructure to distributed cloud computing environments. The shifting is manifested through the rising trends of edge and fog computing environments.

## References

[1] M. A. Rodriguez and R. Buyya, "Scientific Workflow Management System for Clouds," in *Software Architecture for Big Data and the Cloud*, Morgan Kaufmann, 2017, pp. 367-387.

[2] T. Fahringer, R. Prodan, R. Duan, J. Hofer, F. Nadeem, F. Nerieri, S. Podlipnig, J. Qin, M. Siddiqui, H.-L. Truong, A. Villazon and M. Wieczorek, "ASKALON: A Development and Grid Computing Environment for Scientific Workflows," in *Workflows for e-Science: Scientific Workflows for Grids*, London, Springer, 2007, pp. 450-471.

[3] J. Qin and T. Fahringer, Scientific Workflows: Programming, Optimization, and Synthesis with ASKALON and AWDL, Springer, 2014.

[4] P. Blaha, K. Schwarz, G. K. Madsen, D. Kvasnicka and J. Luitz, "WIEN2K, An Augmented Plane Wave+ Local Orbitals Program for Calculating Crystal Properties," Vienna University of Technology, Vienna, 2001.

[5] J. Goecks, A. Nekrutenko and J. Taylor, "Galaxy: A Comprehensive Approach for Supporting Accessible, Reproducible, and Transparent Computational Research in The Life Sciences," *Genome Biology,* vol. 11, no. 8, p. R86, 2010.

[6] C. A. Goble, J. Bhagat, S. Aleksejevs, D. Cruickshank, D. Michaelides, D. Newman, M. Borkum, S. Bechhofer, M. Roos and P. Li, "myExperiment: A Repository and Social Network for the Sharing of Bioinformatics Workflows," *Nucleic Acids Research,* vol. 38, pp. 677-682, 2010.

[7] D. R. Bharti, A. J. Hemrom and A. M. Lynn, "GCAC: Galaxy Workflow System for Predictive Model Building for Virtual Screening," *BMC Bioinformatics,* vol. 19, no. 13, p. 550, 2019.

[8] M. W. C. Thang, X. Y. Chua, G. Price, D. Gorse and M. A. Field, "MetaDEGalaxy: Galaxy Workflow for Differential Abundance Analysis of 16s Metagenomic Data," *F1000Research,* vol. 8, p. 726, 2019.

[9] D. Eisler, D. Fornika, L. C. Tindale, T. Chan, S. Sabaiduc, R. Hickman, C. Chambers, M. Krajden, D. M. Skowronski, A. Jassem and W. Hsiao, "Influenza Classification Suite: An

Automated Galaxy Workflow for Rapid Influenza Sequence Analysis," *Influenza and Other Respiratory Viruses,* 2020.

[10] B. Balis, "HyperFlow: A Model of Computation, Programming Approach and Enactment Engine for Complex Distributed Workflows," *Future Generation Computer Systems,* vol. 55, pp. 147-162, 2016.

[11] M. Malawski, A. Gajek, A. Zima, B. Balis and K. Figiela, "Serverless Execution of Scientific Workflows: Experiments with HyperFlow, AWS Lambda and Google Cloud Functions," *Future Generation Computer Systems,* 2017.

[12] I. Altintas, C. Berkley, E. Jaeger, M. Jones, B. Ludascher and S. Mock, "Kepler: An Extensible System for Design and Execution of Scientific Workflows," in *Proceedings of the 16th International Conference on Scientific and Statistical Database Management*, 2004.

[13] J. Davis, M. Goel, C. Hylands, B. Kienhuis, E. A. Lee, J. Liu, X. Liu, L. Muliadi, S. Neuendorffer and J. Reekie, "Overview of the Ptolemy Project," 1999.

[14] P. Korambath, J. Wang, A. Kumar, J. Davis, R. Graybill, B. Schott and M. Baldea, "A Smart Manufacturing Use Case: Furnace Temperature Balancing in Steam Methane Reforming Process via Kepler Workflows," *Procedia of Computer Science,* vol. 80, pp. 680-689, 2016.

[15] P. C. Yang, S. Purawat, P. U. Ieong, M. T. Jeng, K. R. DeMarco, I. Vorobyov, A. D. McCulloch, I. Altintas, R. E. Amaro and C. E. Clancy, "A Demonstration of Modularity, Reuse, Reproducibility, Portability and Scalability for Modeling and Simulation of Cardiac Electrophysiology Using Kepler Workflows," *PLOS Computational Biology,* vol. 15, no. 3, pp. 1-19, 2019.

[16] E. Deelman, K. Vahi, M. Rynge, R. Mayani, R. daSilva, G. Papadimitriou and M. Livny, "The Evolution of the Pegasus Workflow Management Software," *Computing in Science Engineering,* vol. 21, no. 4, pp. 22-36, 2019.

[17] D. Thain, T. Tannenbaum and M. Livny, "Distributed Computing in Practice: The Condor Experience," *Concurrency - Practice and Experience,* vol. 17, no. 2-4, pp. 323-356, 2005.

[18] E. Deelman, C. Kesselman, G. Mehta, L. Meshkat, L. Pearlman, K. Blackburn, P. Ehrens, A. Lazzarini, R. Williams and S. Koranda, "GriPhyN and LIGO, Building a Virtual Data Grid for Gravitational Wave Scientists," in *High Performance Distributed Computing*, 2002.

[19] K. Wolstencroft, R. Haines, D. Fellows, A. Williams, D. Withers, S. Owen, S. Soiland-Reyes, I. Dunlop, A. Nenadic, P. Fisher, J. Bhagat, K. Belhajjame and F. Bacall, "The Taverna Workflow Suite: Designing and Executing Workflows of Web Services on the Desktop, Web or in the Cloud," *Nucleic Acids Research,* vol. 41, no. 1, pp. 557-561, 2013.

[20] B. B. Misra, "Open-Source Software Tools, Databases, and Resources for Single-Cell and Single-Cell-Type Metabolomics," in *Single Cell Metabolism: Methods and Protocols*, New York, Springer, 2020, pp. 191-217.

[21] R. Tsonaka, M. Signorelli, E. Sabir, A. Seyer, K. Hettne, A. Aartsma-Rus and P. Spitali, "Longitudinal Metabolomic Analysis of Plasma Enables Modeling Disease Progression in Duchenne Muscular Dystrophy Mouse Models," *Human Molecular Genetics,* 2020.

[22] J. Yu and R. Buyya, "Gridbus Workflow Enactment Engine," in *Grid Computing: Infrastructure, Service, and Applications*, CRC Press, 2018.

[23] C. Vecchiola, X. Chu and R. Buyya, "Aneka: A Software Platform for .NET-based Cloud Computing," in *High Speed and Large Scale Scientific Computing*, Amsterdam, IOS Press, 2009, pp. 267-295.

[24] S. Pandey, D. Karunamoorthy and R. Buyya, "Workflow Engine for Clouds," in *Cloud Computing*, John Wiley & Sons, Ltd, 2011, pp. 321-344.

[25] M. H. Hilman, M. A. Rodriguez and R. Buyya, "Task-Based Budget Distribution Strategies for Scientific Workflows with Coarse-Grained Billing Periods in IaaS Clouds," in *Proceedings of the 13th IEEE International Conference on e-Science*, Auckland, 2019.

[26] M. H. Hilman, M. A. Rodriguez and R. Buyya, "Resource-sharing Policy in Multi-tenant Scientific Workflow as a Service Platform," arXiv, 2019.

[27] M. P. Mackley, B. Fletcher, M. Parker, H. Watkins and E. Ormondroyd, "Stakeholder Views on Secondary Findings in Whole-genome and Whole-exome Sequencing: A Systematic Review of Quantitative and Qualitative Studies," *Genetics in Medicine,* vol. 19, no. 3, pp. 283-293, 2017.

[28] D. Dong, Z. Xu, W. Zhong and S. Peng, "Parallelization of Molecular Docking: A Review," *Current Topics in Medicinal Chemistry,* vol. 18, no. 12, pp. 1015-1028, 2018.

[29] J. Kelleher, Y. Wong, A. W. Wohns, C. Fadil, P. K. Albers and G. McVean, "Inferring Whole-genome Histories in Large Population Datasets," *Nature Genetics,* vol. 51, no. 9, pp. 1330-1338, 2019.

[30] M. H. Hilman, M. A. Rodriguez and R. Buyya, "Task Runtime Prediction in Scientific Workflows Using an Online Incremental Learning Approach," in *Proceedings of the 11th IEEE/ACM International Conference on Utility and Cloud Computing*, Zurich, 2018.

[31] A. Gimeno, M. J. Ojeda-Montes, S. Tomás-Hernández, A. Cereto-Massagué, R. Beltrán-Debón, M. Mulero, G. Pujadas and S. Garcia-Vallvé, "The Light and Dark Sides of Virtual Screening: What Is There to Know?," *International Journal of Molecular Sciences,* vol. 20, no. 6, 2019.

[32] C. Grebner, E. Malmerberg, A. Shewmaker, J. Batista, A. Nicholls and J. Sadowski, "Virtual Screening in the Cloud: How Big Is Big Enough?," *Journal of Chemical Information and Modeling,* 2019.

[33] O. Trott and A. J. Olson, "AutoDock Vina: Improving the Speed and Accuracy of Docking with A New Scoring Function, Efficient Optimization, and Multithreading," *Journal of Computational Chemistry,* vol. 31, no. 2, pp. 455-461, 2010.